\documentclass{aa}     
\usepackage{graphicx}

\newcommand{\Mpc}{$h^{-1}$\thinspace Mpc}

\begin{document}   

                       
\title{The mass function of the Las Campanas loose groups of galaxies}

\author{P. Hein\"am\"aki \inst{1,2}, J. Einasto \inst{1}, M. Einasto
 \inst{1}, E. Saar \inst{1}, D. L. Tucker
 \inst{3} and V. M\"uller \inst{4}}

\offprints{P. Hein\"am\"aki}

\authorrunning{P. Hein\"am\"aki  et al.}

\institute{Tartu Observatory, EE-61602 T\~oravere, Estonia,
\and  Tuorla Observatory, V\"ais\"al\"antie 20, Piikki\"o, Finland,
\and Fermi National Acceleration Laboratory, M8 127, P.O. Box 500, Batavia,
IL 60510, U.S.A.,
\and Astrophysical Institute Potsdam, An der Sternwarte 16,
       D-14482 Potsdam, Germany}
\date{Received ; Accepted} 

\titlerunning{The mass function}

\abstract{ We have determined the mass function of loose groups of
galaxies in the Las Campanas Redshift Survey.  Loose groups of
galaxies in the LCRS range in mass from $M \sim 10^{12} {\rm
M}_{\sun}$ to $10^{15} {\rm M}_{\sun}$.  We find that the sample is
almost complete for masses in the interval $5\cdot 10^{13}-8\cdot
10^{14}~{\rm M}_{\sun}$.  Comparison of the observed mass function
with theoretical mass functions obtained from N-body simulations shows
good agreement with a CDM model with the parameters $\Omega_m =
0.3$, $\Omega_{\Lambda} = 0.7$ and the amplitude of perturbations
about $\sigma_8=0.78-0.87$.  
For smaller masses the mass function of LCRS loose
groups flattens out, differing considerably from the group mass
function found by Girardi and Giuricin (2000) and from mass functions
obtained by numerical simulations.

 }

\maketitle

\section{Introduction}

One important constraint of cosmological models is provided by the
mass function of clusters and groups of galaxies.  In the case of
popular $\Lambda$CDM models the amplitude and shape of the mass
function depend primarily on the mean density of matter in the
Universe, $\Omega_m= \Omega_c + \Omega_b$, where $\Omega_c$ and
$\Omega_b$ are the mean densities of the cold dark matter and baryonic
matter in units of the critical cosmological density, respectively.
The amplitude of the mass function also depends on the amplitude of
the power spectrum of density fluctuations, which can be characterised
by the $\sigma_8$ parameter (the linearly extrapolated rms density
fluctuations in a sphere of of 8~\Mpc\  radius).  The local abundance
of rich clusters of galaxies can be used to estimate $\sigma_8$
(Bahcall, Fan \& Cen  \cite{bac:fan}, de Theije, van Kampen \& 
Slijkhuis \cite{the:the}, Cen
\cite{cen:cen}).

Determination of the cluster mass function consists of two tasks:
calculation of cluster masses, and estimation of the spatial density
of clusters.  Masses of clusters can be derived using either X-ray
data and the mass-temperature relation or data from optical surveys
using the velocity dispersion of galaxies in clusters (virial masses).
In a pioneering study by Bahcall and Cen (\cite{bac:cen}; hereafter
BC93), both mass determination methods were used.  Masses were derived
for Abell clusters of all richness classes and for groups of
galaxies. Biviano et al. (\cite{biv:biv}) and
Girardi et al.  (\cite{gir:bor}, hereafter G98) 
calculated virial masses of nearby
clusters. Reiprich \& B\"ohringer (\cite{rei:rei}) used a X-ray
flux-limited sample of galaxy clusters (HIFLUGCS, the HIghest X-ray
FLUx Galaxy Cluster Sample) to obtain a mass function.  The
spatial density of massive clusters of galaxies, according to G98,
exceeds the density found by BC93 by a factor of almost ten.  As shown
by G98, this difference is mainly due to the fact that BC93 assumed a
one-to-one correspondence between the richness of a cluster and its
mass, whereas in reality this relationship has a large scatter. 
A new determination of the mass function of
loose groups of galaxies has been provided by Girardi and Giuricin
(\cite{gir:giu}). 

The sample of loose groups of galaxies identified in the Las Campanas
Redshift Survey (LCRS) by Tucker et al.  (\cite{tuc:tuc}, hereafter
TUC) provides a possibility to derive a new independent estimate of the
mass function of clusters and groups of galaxies.  Loose groups
represent the most numerous component of galaxy clustering, and so we
should be able to determine their statistical properties much better
than those for rare rich clusters of galaxies.  The virial masses of loose
groups of galaxies in the LCRS range from $M \sim 10^{12} {\rm
M}_{\sun}$ to $10^{15} {\rm M}_{\sun}$; these masses were estimated by
TUC using velocity dispersions of galaxies and harmonic radii of
groups.

In this paper we shall estimate the mass function of LCRS loose
groups.  Our main task is to derive the spatial density of groups of
various masses.  The volume of the LCRS sample is well determined, thus
we hope to get an unbiased estimate of the spatial density of groups.
We shall compare the empirical mass function with theoretical
mass functions found using numerical simulations of structure
evolution.  
In our simulations we assign to a halo all
the particles identified as members of the halo by halo finder. 
In observations this means that the velocity dispersion of galaxies
is assumed to be identical to that of the dark matter.  This
comparison allows us to check the consistency of structure evolution
models with empirical data and to find the set of cosmological
parameters which brings models into agreement with data.  We also 
determine the mass interval where the sample of groups is not
influenced by selection effects.

\section{The Data}  

\subsection{Observations}

The LCRS (Shectman et al. \cite{she:she}) is an optically selected
galaxy redshift survey where a multi--object spectrograph was used to
measure simultaneously redshifts of 50 or 112 galaxies.  Extending to
a redshift of $z\approx 0.2$, the catalogue covers 6 slices of size on
average $1.5^{\circ} \times 80^{\circ}$, containing 23,967 galaxies
with measured redshifts within the official survey photometric and
geometric limits.  Three slices are located in the northern Galactic
hemisphere, centred at declinations $\delta= -3^{\circ},~ -6^{\circ},~
-12^{\circ}$, and the other three slices are located in the southern
Galactic hemisphere, centred at declinations $\delta= -39^{\circ},~
-42^{\circ},~ -45^{\circ}$.  TUC applied a friends-of-friends (FoF)
percolation algorithm to extract the catalogue of Las Campanas loose
groups of galaxies (hereafter LCLG).  The linking parameter was chosen
to get a density enhancement limit of $\delta n/n\ge 80$. The minimum
group membership was chosen to be three.

Because the spectroscopy was carried out for each field either via a
50 or a 112 fibre multiobject spectrograph, the selection criteria
varied from field to field.  The nominal apparent magnitude limits for
50 fibre fields were $16.0 \le R \le 17.3$ and for 112 fibre fields
the limits were $15.0 \le R \le 17.7$.  According to TUC great effort
was put forth in accounting for these field-to-field sampling
variations.  The general properties of the 50 fibre and the 112 fibre
groups agree well with properties of groups found from other surveys,
thus we shall use the whole TUC group sample.

Only groups with redshifts $10,000 \le cz \le 45,000$ km~s$^{-1}$
were included into the LCLG sample.  TUC concluded that the LCLG is a
useful sample for a variety of studies requiring an unbiased
collection of loose groups.  It is based on the LCRS galaxy sample, which
is the first redshift survey that can claim to enclose a reasonably
fair sample of the nearby universe.

The complete LCLG list includes 1495 groups. TUC also introduced a
``clean sample'', where groups with four potential bias factors are
excluded: 1) groups which are too close to a slice edge, 2) groups
which have the crossing time greater than the Hubble time, 3) groups
with the corrected velocity dispersion zero what can happen 
since TUC subtracted an redshift measurement error of 67 km/s 
in quadrature from each group velocity dispersion, and 4) groups containing
a 55 arc-sec orphan galaxy, i.e. a galaxy with no measured
redshift. The last bias was caused by technical reasons (the fibre
separation limit, which prevents the observation of neighbouring
galaxies within 55 arc-seconds of each other).  In the full sample
this effect was compensated for by reintroducing lost galaxies,
assigning to them a redshift equal to that of its nearest neighbour,
convolved with a Gaussian of width $\sigma=200$~km~s$^{-1}$.

\subsection{Simulations}

A low density CDM universe with a cosmological constant ($\Lambda$CDM)
is widely regarded as the best model compatible with most of the
currently available data; e.g. with the microwave background
anisotropy measured by BOOMERANG (de Bernardis et al. \cite{ber:ber})
and MAXIMA I (Hanany et al. \cite{han:han}) experiments, and with data on the
large scale structure of the universe: two-point correlation functions and
power spectra, and the spatial density of mass-limited samples of
galaxy clusters (Governato et al. \cite{gov:gov}, Colberg et
al. \cite{col:col}, Pierpaoli et al. \cite{pie:sco}).

For the present study we employ a flat cosmological model
($\Omega_m+\Omega_{\Lambda}=1$) with the following parameters:
the matter density $\Omega_m=0.3$, the
baryonic density $\Omega_b=0.04$, the vacuum energy density
(cosmological constant) $\Omega_{\Lambda}=0.7$, and the Hubble
constant $h=0.7$ (here and throughout this paper $h$ is the
present-day Hubble constant in units of 100 km s$^{-1}$ Mpc$^{-1}$).
The simulations were performed using a P3M code (Couchman
\cite{couch91}) for a cube of 200~\Mpc\ size and a $256^3$ mesh and
for the same number of particles; each particle has a mass of
$4.0\times 10^{10} h^{-1} {\rm M}_{\sun}$.  The transfer function was
computed using the CMBFAST code by Seljak and Zaldarriaga
(\cite{sel:zal}). The rms mass density fluctuation parameter of this
model is $\sigma_8=0.87$.

We also calculated an additional model using the same code with $128^3$
particles in a cube of size $L=100$~\Mpc, and the cosmological
parameters $\Omega_m=0.3$, $\Omega_{\Lambda} = 0.7$, $\Omega_b=0.05$,
and $h=0.65$; this COBE normalised model has the density fluctuation
parameter $\sigma_8 = 0.78$. 

We used the FoF algorithm to identify CDM halos.  The only free
parameter in the FoF method is the linking length, which is defined as
the maximum separation between particles which are still joined into
groups.  In our case the linking length was chosen as 0.23 in units of
the mean particle separation, which approximately selects the matter
density contrast $\delta n/n= 80$.  This density contrast was the one
chosen by TUC to extract the group catalogue from LCRS, and it is
typical of the values used in extracting groups from observational
galaxy catalogues.  While in the Einstein-de Sitter universe the
overdensity within the virial radius of a cluster is $\delta n/n=
178$, in a low density spatially flat universe with a cosmological
constant it is $\delta n/n\approx 178\Omega_m^{-0.6}$ (White et
al. \cite{whi:efs}).  The corresponding overdensity for our simulation
is $\delta n/n \approx 366$.  This means that the group sample
determined using a low density contrast ($\delta n/n= 80$) contains
groups which can be in an uncertain dynamical state and do not have to
be virialised. When particles outside the virialised core are included
in the groups, the masses of simulated groups may be overestimated.
For comparison we also calculated the mass function  
of publicly available numerical
simulations of the Virgo consortium\footnote{{\tt
http://www.MPA-Garching.MPG.DE/Virgo/}}, where Jenkins et
al. \cite{jen:fre} selected FoF-groups with a
linking length of 0.2 of the mean inter-particle separation.

\section{Results}

\subsection{Selection effects in the LCLG sample}

At first we discuss selection effects and show how we took 
them into account when calculating the mass function.
There are two main selection effects in the LCRS: 1) observations are
performed in a fixed apparent magnitude interval which transforms to a
distance dependent absolute magnitude 
interval, galaxies fainter or brighter than this interval are not
included in the survey; 2) depending on the field 50 or 112 galaxies
were measured for redshifts, the actual number of galaxies in the
magnitude window could be larger, thus the sample is diluted.

\begin{figure}
\resizebox{\hsize}{!}{\includegraphics{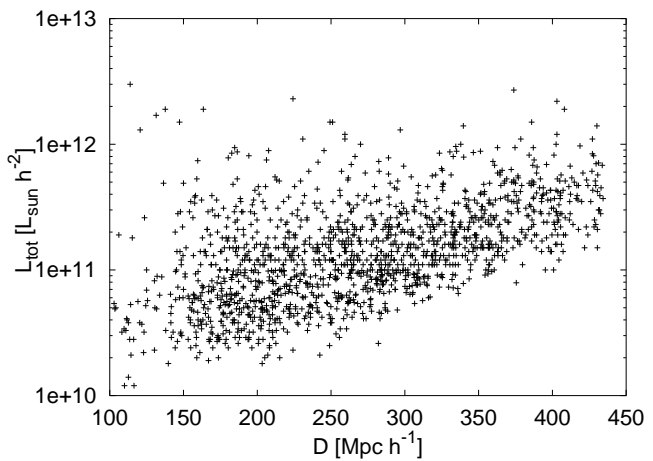}}
\caption{The total luminosity of the Las Campanas loose groups
as a function of their distance} 
\label{lumin}
\end{figure}

\begin{figure}
\resizebox{\hsize}{!}{\includegraphics{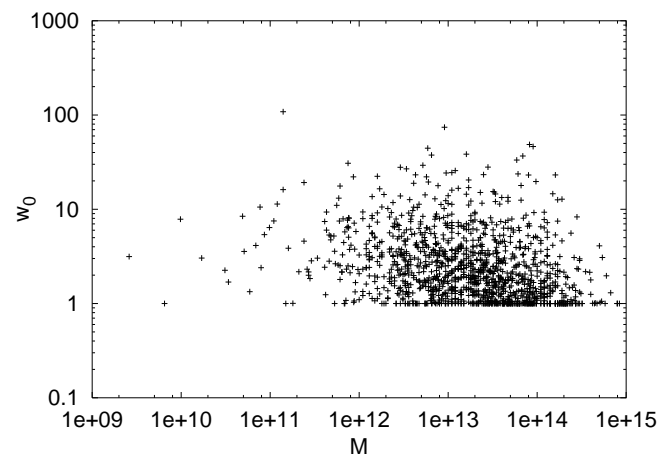}}
\caption{The weighting parameter $w_0$ as a function of 
the group mass} 
\label{painot}
\end{figure}

The first selection effect makes it impossible for galaxies outside
the window to enter the survey.  Consequently groups consisting of
faint galaxies occur only in the nearest region of the survey; with
increasing distance fainter groups gradually disappear from the
sample.  This effect is seen in Figure~\ref{lumin} where the total
luminosity of LCRS loose groups is shown as a function of distance.
This selection effect can be statistically corrected using individual 
weights in
calculation of the group mass function, following the procedure by
Moore, Frenk \& White (\cite{moo:moo}) and 
Girardi \& Guiricin (\cite{gir:giu}).  
Namely, we weight each group by
$w=1/\Gamma$, where $\Gamma$ is the volume, which corresponds to the
absolute magnitude limit of the third brightest galaxy of a group; if 
$\Gamma > \Gamma_0$ (the full volume of a slice), we 
take $\Gamma = \Gamma_0$.
The data to calculate these limits have been 
tabulated by TUC separately for each of the 327 spectroscopic fields
of the LCRS.  Figure ~\ref{painot} shows the distribution of the
normalised weights $w_0$ (in units of the full volume of a slice)
as a function of group mass. We see that there is a weak relation
between cluster mass and the relative weight: massive groups contain
as a rule sufficiently bright galaxies and can be observed in the
whole volume of the survey, thus for massive clusters the relative
weight is closer to unity.

Now we consider the second selection effect.  To estimate the degree
of dilution of the LCRS we compare the surface density of galaxies of the
LCRS with the surface density of the Sloan Digital Sky Survey. Here
the goal was to measure all galaxies in the magnitude window. The
number of galaxies with measured redshifts in the Early Data Release
of SDSS is 70 per square degree, while the LCRS has only 20 galaxies per square
degree in the $-6^{\circ}$ slice, and $30 - 38$ per square degree in
the rest of slices. The faint end limit in most slices 
of the LCRS is almost same as for the 
SDSS ($-17.7$), thus the difference in number density
is due to dilution of the LCRS.  Most slices of the LCRS are
diluted by a factor of about 2, and the slice $-6^{\circ}$ by a
factor of 3.5.  Dilution decreases not only the number of galaxies but
also the number of clusters and groups, as the number of galaxies in
a group may fall below 3 and the group will not be included
in the catalogue. The number of groups
in the  $-6^{\circ}$ slice is actually lower than in other slices 
by a factor of about 2.  

However, uniform dilution does not influence strongly
the mass function of groups. Firstly, virial
masses of groups practically do not change. When determining
virial masses, group members play the role of test
particles, which move in a common gravitational
potential. Using less test particles gives the same 
estimate for the potential and for the virial mass
as before, only
with a larger variance. We checked this conclusion
by diluting simulated group catalogues.

	Dilution may affect the group catalogue
only by reducing the group richness of smaller
groups below the catalogue limit. These groups will
drop out of the catalogue and the number density
of smaller (less massive) groups will fall.
We estimated this correction, using group
catalogues obtained in our numerical simulations
and randomly diluting the groups by 50\%, as estimated above.
We found that the conditional probability distribution
$P(\log N|M)$ ($N$ is the richness of a diluted group, $M$ is the
virial mass of the group, and we use decimal logarithms here)
is close to a Gaussian with a mean 
\[
\overline{\log N(M)}\approx\log(M/{\rm M}_{\sun}) -11.5
\]
and a rms error $\sigma$ that is also a function of the
virial mass, 
\[
\sigma(M)\approx 0.3(14-\log(M/{\rm M}_{\sun})). 
\]
The fraction $\alpha$
of the groups, which remain in the catalogue after dilution,
is given then by the error integral:
\[
\alpha(M)=0.5\mbox{erfc}\left(\frac{\log N_L-\overline{\log N(M)}}
	{\sqrt{2}\sigma(M)}\right),
\]
where $N_L$ is the catalogue limit (3 and 5, in our case) and 
$\overline{\log N(M)}$ and $\sigma(M)$ are given above.
In order to restore the original differential mass function,
we have to multiply it by the factor $\alpha^{-1}$.
This dilution correction is larger than 1 only in the mass interval
$10^{12}$--$10^{13}{\rm M}_{\sun}$, reaching 1.95 at its lowest end
(for $N_L=3$) and 2.67 for $N_L=5$. As the differential mass
function is small in this mass interval, this correction
does not change appreciably the final integral mass function.
Also, the dilution correction does not change the high mass end of the
mass function at all.

\subsection{Mass function of LCLGs}

The mass function (MF) of galaxy clusters/groups is defined as the number
density of clusters above a given mass $M$, \mbox{$n(>M)$}.  To
construct the group mass function from a group sample one
needs accurate group masses and well defined volume and selection
functions of the sample.  We have used the masses estimated by TUC,
who assumed that the groups were virialised and
calculated virial masses of groups as:
\begin{equation} 
{\rm M}_{vir}=\frac{d\sigma_{los}^2R_h}{G},
\label{vm1}
\end{equation} 
where $G$ is the gravitational constant, $R_h$ is the harmonic radius of
the group, $\sigma_{los}$ is the group line-of-sight velocity dispersion,
and $d=6$ in the case of a spherically 
symmetric velocity distribution of the group.

We calculated the volumes of slices as follows:
\begin{equation} 
V= {\cos \delta_m \Delta\alpha \Delta\delta \over 3} (r^3_2 - r^3_1); 
\label{vm2}
\end{equation} 
here $\delta_m$ is the mean declination of the slice, $\Delta\delta$
and $\Delta\alpha$ are the widths of the slice in declination and right
ascension (in radians), and $r_1$ and $r_2$ are the lower and upper
distance used in calculations. This formula is valid for spatially flat (k=0) 
cosmologies.

  \begin{table}
      \caption[]{The  numbers of groups 
in the LCLG catalogue. Columns give the mean declination of a slice,  
the number of all groups by TUC, the number of groups in the clean sample 
and the volume of the slice.} 

         \label{KapSou}
      \[
         \begin{tabular}{lcccc}
            \hline
            \noalign{\smallskip}
            Slice $\delta$ &TUC&Clean & Volume [$Mpc^3/h^3$]\\
            \noalign{\smallskip}
            \hline
            \noalign{\smallskip}
$-3^{\circ}$ &288  &80  &  927902  \\
$-6^{\circ}$ &147  &37  &  823452  \\
$-12^{\circ}$&276  &73  &  874202  \\
$-39^{\circ}$&249  &71  &  971817  \\
$-42^{\circ}$&257  &69  &  988435  \\
$-45^{\circ}$&256  &63  &  941531  \\
Total        &1473 &393 & 5527339  \\
Mean         &     &    &  921223  \\
            \noalign{\smallskip}
            \hline
         \end{tabular}
      \]
   \end{table}

The comoving cosmological distance   
$r$ is a function of measured redshift and depends on the 
cosmological model as (e.g. Peebles \cite{peebles}): 
\begin{equation}
r = \frac{c}{H_0}\int_0^z\frac{dz'}{E(z')},
\end{equation}
where the function $E(z)$ is given by
\begin{equation}
E^2(z)=\Omega_m(1+z)^3+(1-\Omega_m-\Omega_\Lambda)(1+z)^2+\Omega_\Lambda.
\end{equation}

The LCLG survey extends up to the redshift $z \sim 0.15$.  Tucker et
al. (\cite{tuc:tuc}) used the Einstein-de Sitter cosmology (assuming
$\Omega_m=1, \Omega_\Lambda=0$) to determine distances in the
LCLG survey, while modern data prefer the parameters $\Omega_m$=0.3 and
$\Omega_\Lambda$=0.7. Thus for the present study we recalculated
comoving distances for all galaxies of the LCLG survey. For the
particular redshift range $10000 \le cz \le 45000$ km~s$^{-1}$ the
ratio between the comoving volumes in these two models is about 0.8.
The numbers of groups for all slices, for the clean sample and the volume
of each slice in the LCLG catalogue are shown in Table 1. Only
groups with defined mass are included (for some groups the velocity
dispersion, corrected for random errors, is zero and mass
determination is impossible).

\begin{figure}
\resizebox{\hsize}{!}{\includegraphics{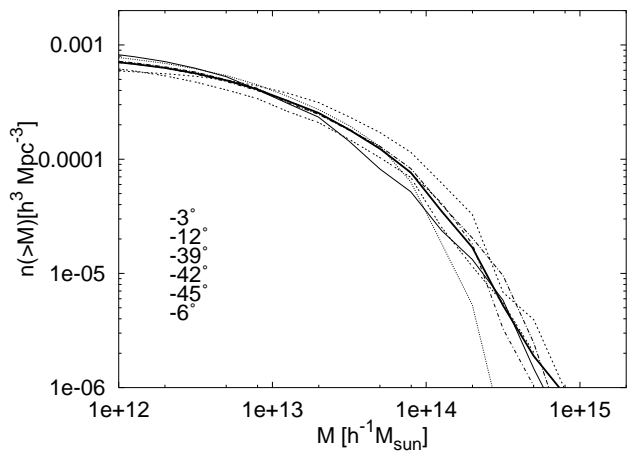}}
\caption{The cumulative mass functions for all six slices 
in the LCLG. The thin  
curves from top to bottom are labelled as seen at 
the low mass end. The thick solid curve shows the average mass function for  
all the clusters. These mass functions are not corrected for dilution.} 
\label{mfLCslice}
\end{figure}

The MF was constructed by sorting the masses in logarithmic bins of
width 0.2. Errors in group masses (an estimate of the rms error in
virial mass has been found by TUC) were accounted for 
by convolving group masses with a 
log-normal distribution of the same mode and variance.  
Figure ~\ref{mfLCslice} shows the mass
function of LCLG for all six slices.  To have a general picture we
added slices together and found the average mass function, 
denoted by a thick solid line.  We found that the
average mass function and the mass function of the clean sample
distributions are very similar, which indicates that the differences between 
the clean sample and the full sample are unimportant for
the MF.  We also tried excluding the slice $\delta= -6^{\circ}$ and
found that it has only a weak influence on the total MF.
In order to better show the differences between various slices
we did not correct these mass functions for dilution.
The dilution-corrected final mass function is given in Table~\ref{KappSou}
and in Fig.~\ref{massf}.

  \begin{table}
      \caption[]{The cumulative mass function of LCLG}

         \label{KappSou}
      \[
         \begin{tabular}{lc}
            \hline
            \noalign{\smallskip}
  log $M$ [$h^{-1}{\rm M}_{\sun}$] & Density [$h^{3}Mpc^{-3}$] \\ 
            \noalign{\smallskip}
            \hline
            \noalign{\smallskip}

1.26E+12 & 7.56E-04 \\
3.16E+12 & 5.78E-04 \\
7.94E+12 & 4.09E-04 \\
1.26E+13 & 3.23E-04 \\
3.16E+13 & 1.81E-04 \\ 
7.94E+13 & 7.63E-05 \\
1.26E+14 & 3.51E-05 \\
3.16E+14 & 5.22E-06 \\
7.94E+14 & 8.77E-07 \\

            \noalign{\smallskip}
            \hline
         \end{tabular}
      \]
   \end{table}

There is a clear difference in the MF between slices.  The two 
northern slices at $\delta= -6^{\circ}$ and $\delta= -12^{\circ}$ 
differ from others, having the lowest and the highest
amplitudes at the faint end of the MF. 
We suppose that this discrepancy
between mass functions are due to differences in dilution, 
since the slice $\delta= -6^{\circ}$ was observed by the 50 fibre
spectrograph only and the $\delta= -12^{\circ}$ slice by the 112 fibre
spectrograph only. This discrepancy cannot be removed by a 
luminosity-weighting procedure.  The remaining differences are
probably due to cosmic variance.  
The slices of the LCRS survey intersect large voids and
superclusters.  In such slices, loose groups form a variety of
structures that may influence the observed mass function at its
massive end, since the most massive groups (clusters) are associated
with supercluster structures.

Figure~\ref{mfLCslice} shows clearly the flattening of mass
functions at the low mass end.  Such flattening is usually interpreted
as the result of incompleteness of a sample (Girardi et al.
\cite{gir:bor}), although it could also be real, describing a
diminishing number of groups of smaller masses.  Thus, we have to
study the completeness problem in more detail.

\begin{figure}
\resizebox{\hsize}{!}{\includegraphics{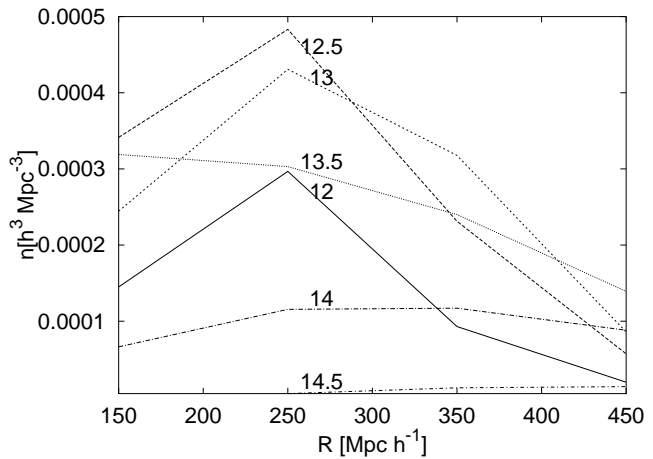}}
\caption{The volume-density of groups as a function of 
distance. Each line represents a different mass interval and is
labeled by the decimal logarithm of its mean $M/{\rm M}_{\sun}$.}
 \label{figvdf}
\end{figure}

To estimate the mass completeness limit of the LCLG we can study the
volume-density of the groups (within different mass intervals) as a
function of distance. In Figure~\ref{figvdf} lines denote mean values
of the density in various mass ranges. 
Any rapid fall of the lines indicates incompleteness
for that mass interval. We find from the
Figure~\ref{figvdf} that below masses of $10^{13.5}-10^{14}{\rm M}_{\sun}$
our sample is probably incomplete.

Another way 
to estimate the mass completeness limit of the LCLG is to use the
velocity dispersion distribution function, VDF (Fadda et al.
\cite{fed:fed}). The VDF is defined similarly to the MF, but for the
velocity dispersion of groups. 
Assuming
that the VDF can be described by a simple power law function (note
that this assumption does not have a clear physical justification),
Fadda et al.  (\cite{fed:fed}) set the completeness limit for their
sample at the point where the power-law exponent of the VDF starts to
change. For the LCLG this point is reached at the line-of-sight
velocity dispersion $\sigma_{los} \approx 250$ km/s.

\begin{figure}
\resizebox{\hsize}{!}{\includegraphics{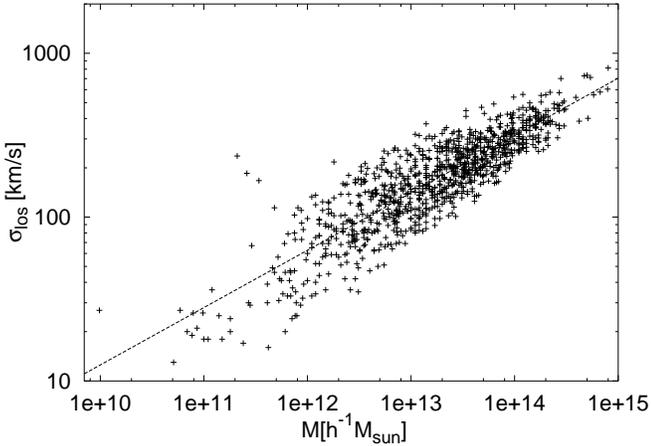}}
\caption{The velocity dispersion as a function 
of the mass for the LCLG catalogue. The line shows the analytical 
(least square) fit.}
\label{figvm}
\end{figure}

Figure~\ref{figvm} shows the relation between $\sigma_{los}$ and mass
$M$ for the LCLG catalogue. Equation (\ref{vm1})  
defines the relation between these
quantities via the harmonic radius of the group. The dotted line
illustrates a power-law fit:
\begin{equation} 
\sigma_{los}=a{\left( hM \over 10^{15} {\rm M}_{\sun}\right)}^c ~{\rm km/s},
\label{sigma-m} 
\end{equation} 
where $a=819$ and $c=0.35$.  The result is in quite good
agreement with previous results by del Popolo \& Gambera  (\cite{pop:popgam})
and del Popolo et al.  (\cite{pop:popetal}), 
$a=842~{\rm km/s}$ and $c=0.33$, based on
N-body simulations and X-ray observations. A value 1/3 for the
exponent $b$ means that the system is virialised.  The fact that the
LCLG follow rather closely the virial relation shows that the possible
contamination is small and the groups are physical.  Setting the
completeness limit to $\sigma_{los} = 250$ km/s, based on the VDF, in
equation (\ref{sigma-m}), we get an estimate for the 
completeness limit in mass between $10^{13} \le M \le 10^{14}
h^{-1}{\rm M}_{\sun}$, with the mean value about $M\sim 5\times 10^{13}
h^{-1}{\rm M}_{\sun}$. This result agrees with the result of the first
analysis.

\begin{figure*}
\centering
\resizebox{.45\hsize}{!}{\includegraphics{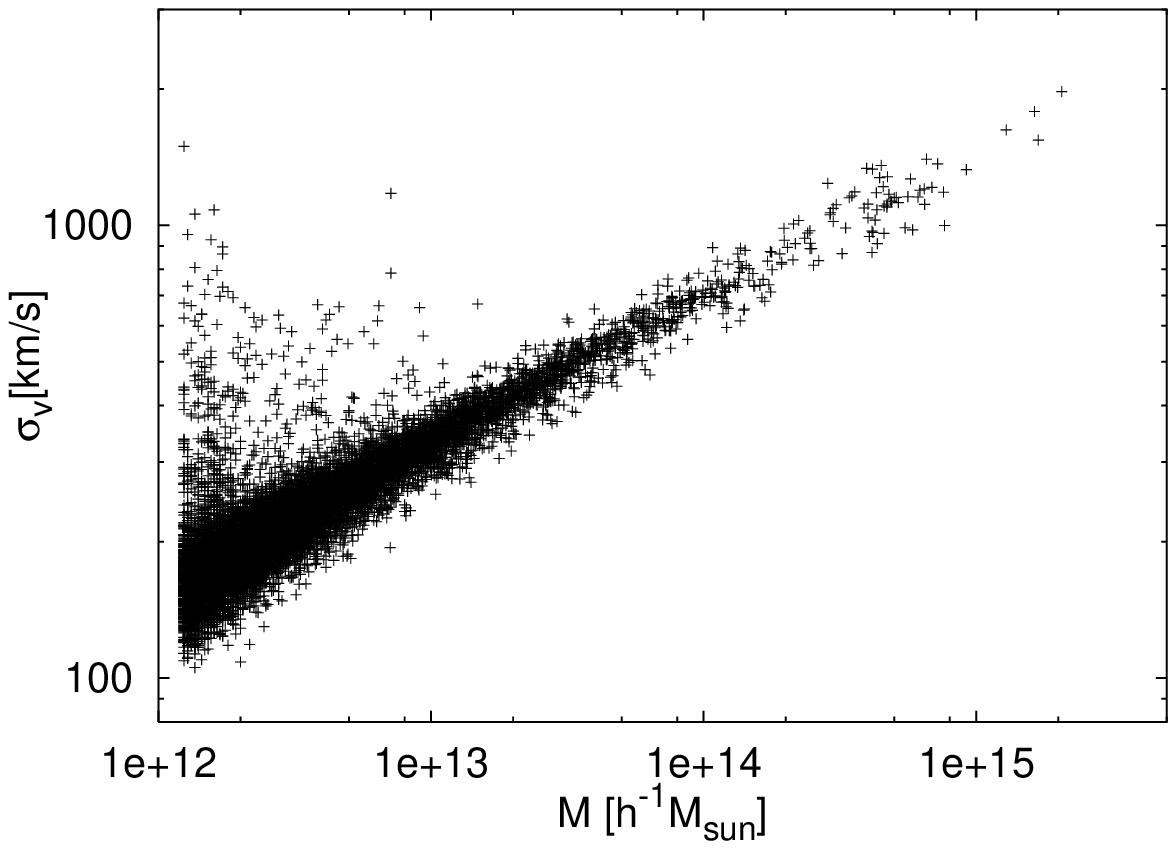}}
\resizebox{.45\hsize}{!}{\includegraphics{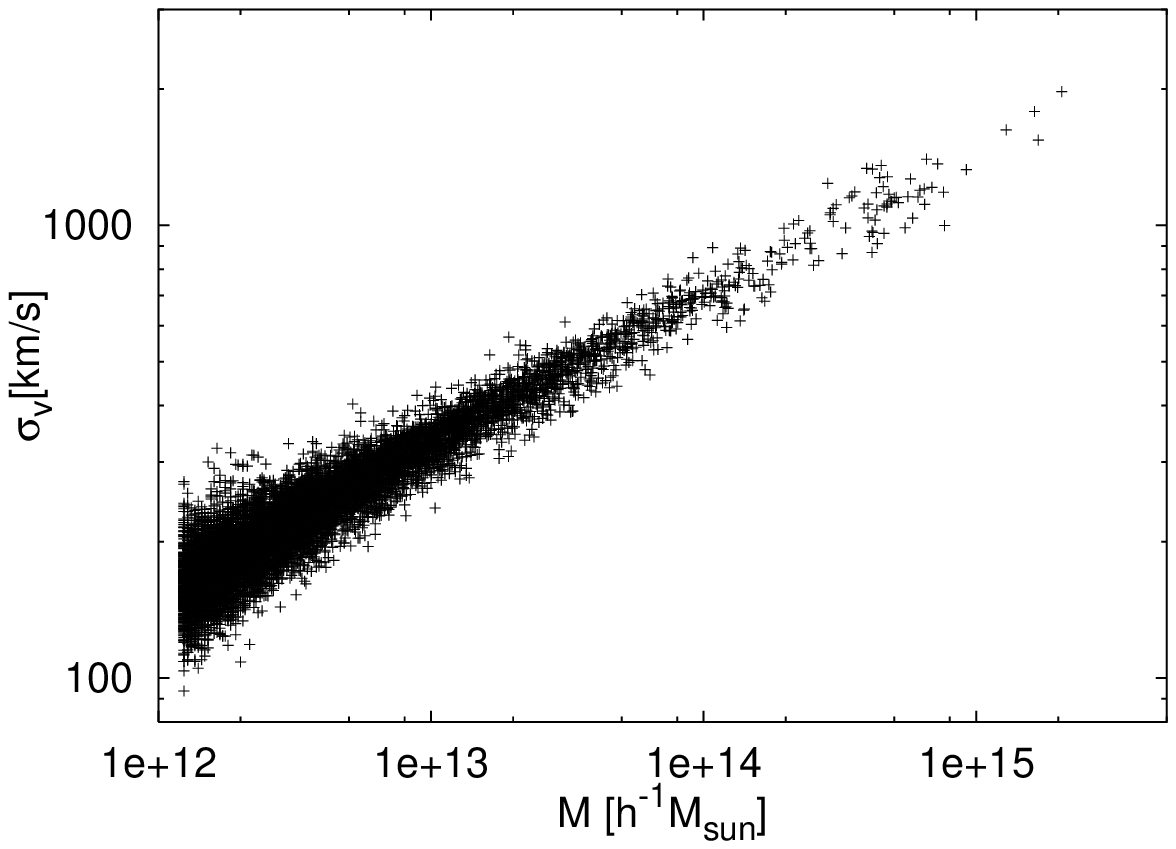}}
\caption{The left panel shows the velocity dispersion of dark matter
particles in halos as a function of the halo mass. The one-to-one
correspondence between the number of particles and the halo mass 
is adopted. The right panel gives the same relation,  but
unbound groups are removed.}
\label{nv}
\end{figure*}

\subsection{Reliability of groups} 

Ramella et al. (\cite{ram}) and Girardi \& Giuricin (\cite{gir:giu})
noted that groups with a few (less than 5 members)
may constitute not real, but pseudo-groups. On average, in pseudo-groups 
the velocity dispersion is considerably larger than in real groups
(Ramella et al. \cite{ram}).  Indeed in the LCLG sample, groups with
less than 5 galaxies have a very large scatter of velocity
dispersions, and some velocity dispersions are extremely large. Analysis of 
a subsample, where poor groups with only 3 or 4 members are excluded, 
shows that in
the $w - M$ plane (as in Figure~\ref{painot}) the scatter 
for this subsample is somewhat
smaller than for the whole sample. However, exclusion of all poor
groups decreases the mass function considerably in the whole mass
range (see Figure ~\ref{massf}, where the mass function is given for
two cases, for all groups and for groups with the number of members
($N_m \ge 5$).  This 
indicates that at least part of small groups with a large velocity
dispersion (and thus a large mass) are probably pseudo-groups  
(studies by Einasto et al. \cite{{ein02}} indicate 
that such groups are located mostly in low density regions). 
On the
other hand, most of small groups have velocity dispersions in the same
range as richer groups, and are 
probably real. In other words, exclusion of all small groups
introduces an additional selection effect, which is difficult to
quantify.

\section{Discussion}

\subsection{Comparison with numerical simulations}

Now we compare our mass function with mass functions found for
groups in numerical simulations.  
Figure~\ref{nv} shows the velocity dispersion of dark matter particles
in groups as a function of the halo (group) mass in our numerical simulations.  We
see that in small groups with mass less than $\sim 10^{13}
h^{-1}{\rm M}_{\sun}$ (which contain less than $\sim 250$ particles) the
scatter of the velocity dispersion is very large. This phenomenon can
be explained if we assume that small groups are not yet virialised and
are contaminated by non-virialised particles, or alternatively are
composed of several small subgroups moving fast relative to each
other.  Another explanation for this effect is that it is a result of
the chosen $\delta n/n$ --- the FoF algorithm collects particles from
outskirts of small clusters although they do not belong dynamically to
the group.  The fraction of interlopers in simulated loose groups can
become rather high, about 20--30\%, as found by Diaferio et
al. (\cite{diaferio}).

To test the second assumption we built groups, using a lower
linking length $b=0.17$ (in units of the mean particle separation).
This did not change considerably the scatter of the velocity dispersion
among the groups with less than 250 particles.  Thus the first explanation
is more likely.

To eliminate the effect of unbound groups in simulations we calculated
both the potential and kinetic energy for all groups and excluded
unbound particles, using the condition $|E_{pot}| < 0.8 E_{kin}$
($E_{pot}$ is the potential energy and $E_{kin}$ the kinetic energy of
a group) for an unbound particle.  We split groups with a high
kinetic energy as compared to the potential energy to separate
interacting or merging groups.  To this aim we employed a 6D-group
finder with a linking measure $\Delta r^2/r^2 + \Delta
v^2/\sigma_v^2$, where $r$ and $\sigma_v$ are the group effective
(half-mass) radius and the velocity dispersion.
This procedure affected only about 10\% of the halos, with many of the
halos having a small number of DM particles.  We show the results of our 
clean-up procedure in
the right panel of Figure~\ref{nv}, which is similar to the left panel,
only non-virialised particles and bypassing groups are excluded. We
see that our procedure to eliminate non-virialised particles works
well --- the low-mass groups follow the same trend as observed for
the high-mass groups and clusters.

We have calculated the mass functions for the sample of all
groups and for the sample of virialised groups in our simulations.  
These functions are very similar, only 
for groups of very small mass ($M < 10^{12}~h^{-1}{\rm
M}_{\sun}$) the MF of virialised groups differs from that of all
groups by less than 1\%. Thus the contribution of unbound groups to
the total mass function is practically negligible.

\begin{figure}
\resizebox{\hsize}{!}{\includegraphics{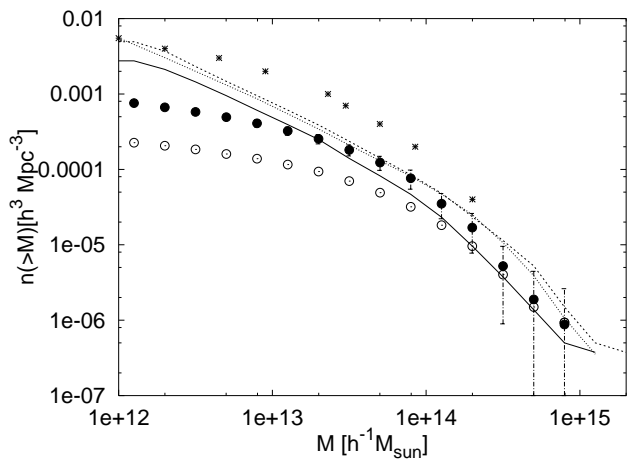}}
\caption{Comparison of the dilution-corrected observed mass functions
and the theoretical mass functions.
The observed mass functions are shown with symbols: 
filled circles (the LCLG whole sample), 
open circles (the LCLG $N_m \ge 5$) and stars 
(Girardi \& Giuricin (\cite{gir:giu}). The dashed line
shows our $256^3$ simulation MF
($\sigma_8=0.87$), the solid line shows our $128^3$ simulation  
($\sigma_8=0.78$) MF, and the dotted line shows the Virgo Consortium
simulation, MF. For clarity Poissonian errors 
are shown only for one case (the full mass function).}  
\label{massf}
\end{figure}

Figure ~\ref{massf} 
shows the mass function of dark matter halos in
our simulation together with the MF of the LCLG.
The Poisson error bars (square root of counts per bin 
multiplied with the mean weight) 
for the full LCLG mass function are shown by the dash-dotted
lines. In order not to overcrowd the figure, we have not shown
the error bars for the $N>=5$ sample; they are very close
to those of the full sample. For small masses 
($M<5 \cdot 10^{13}{\rm M}_{\sun}$)
the error bars are smaller than the size of the plotting symbols.
For comparison, Figure
~\ref{massf} shows also the mass function of the simulations by the Virgo
Consortium, $\Lambda$CDM-gif (Jenkins et al. \cite{jen:fre},
Kauffmann et al. \cite{kau:col}).  In the Virgo simulation the side of
the box was 141.3~\Mpc, the matter density $\Omega_m=0.3$, the
cosmological term $\Omega_{\Lambda} =0.7$, the particle mass
$1.4\times 10^{10} {\rm M}_{\sun}/h$, and the number of the particles
$256^3$.  Figure ~\ref{massf} shows that our $128^3$ simulation with the
amplitude parameter $\sigma_8=0.78$ gives the best fit among the models 
we used to simulate the LCLG
subsample at the massive end ($2\cdot10^{14}{\rm M}_{\sun}-
10^{15}{\rm M}_{\sun}$)
of the MF. For smaller masses, down to $4\cdot10^{13}{\rm M}_{\sun}$
the model with a higher $\sigma_8=0.87$ gives a better fit.
The cluster MFs in the Virgo Consortium simulation and in our $256^3$
simulation are quite similar. 
For yet smaller masses, where we suspect that the LCLG sample
is incomplete, the LCLG mass function flattens out, in contrast to the rising
simulated mass functions.

Most of 
earlier results give $\sigma_8 \sim 0.9 - 1$ for the $\Omega_m=0.3$
scenario (Eke et al. \cite{eke}, Viana and Liddle \cite{viena}
etc.). This value agrees with the recent result by Pierpaoli et
al. (\cite{pie:sco}), $\sigma_8=1.02$.  Lower values are obtained 
based
on X-ray cluster samples. Reiprich \& B\"ohringer (\cite{rei:rei})
derived for $\Omega_m=0.3$, $\sigma_8 \sim 0.68$.  From the ROSAT 
Deep Cluster Survey Borgani et al. (\cite{borgani}) derived 
$\Omega_m=0.35$ and  $\sigma_8 \sim 0.66$.  
Recently Lahav et al. (\cite{lah}) obtained $\sigma_8=0.73$ using 
the 2dFGRS+CMB data. Finally, Bahcall et al. (\cite{bah}) 
used the SDSS data and got  
$\sigma_8=0.72$ for $\Omega_m=0.3$. 
In principle, $\sigma_8$ can be determined on the 
basis of the CMB data only; future experiments as MAP and PLANCK will 
help to pinpoint this parameter.

\subsection{Comparison with earlier work}

The mass function determined by
Girardi \& Giuricin (\cite{gir:giu}) is 
shown in Figure ~\ref{massf}. This group catalogue was constructed using 
a subsample of galaxies of the Lyon-Meudon 
Extragalactic Database. Groups were identified 
using two different methods: the FoF algorithm (P-groups, 453 groups) and 
the hierarchical method (H-groups, 498 groups). 
These group samples extend until $cz \le 5500$ km~s$^{-1}$.    

Here we used  the P-data from figure 3 in 
Girardi \& Giuricin (\cite{gir:giu}). In that case they chose
the subsample of the nearby groups with $cz \le 2000$ km~s$^{-1}$.    , 
which they assumed to be a complete sample and a good representation 
of the total population. However, this sample covers only the local 
supercluster, while the LCLG catalogue
extends up to distances $\sim$ 450~\Mpc. This 
volume contains already a variety of structures. 

We see that in the high mass range ($M \ge 10^{14} h^{-1}{\rm
M}_{\sun}$) the MFs are rather similar. Around
masses $M  \approx 5\cdot10^{13}$ the mass functions of the simulations and
the LCLG start to differ from each other.  
Also, at these lower masses the
mass function by Girardi \& Giuricin (\cite{gir:giu}) is considerably
higher than the LCLG mass function.

This is an intriguing result that certainly needs further analysis.
We see that although loose groups promise to provide better
statistics, there are a number of unsolved problems that do not
permit a realisation of this promise at the moment.  There
are two problems in the LCRS group sample: reliable determination
of masses of small groups, and selection effects.  Virial mass
estimates can be in error; Girardi and Giuricin (\cite{gir:giu}) try
to correct for this. Their correction is rather complicated and could
induce its own errors.

The simplest explanation of the difference between the mass functions
found in the present paper and that of Girardi and Giuricin
(\cite{gir:giu}) is a mass-dependent selection effect. The LCRS is a
strongly diluted sample of galaxies,  and it is biased against
low surface brightness galaxies. 
However, as shown above (Sec. 3.2), dilution
practically does not affect the integral mass function.

The second effect, the absence of low-brightness galaxies 
is probably much less present in the Lyon-Meudon galaxy catalogue, from
which Garcia (\cite{garcia})  
extracted the catalogue of local groups that served as a source
for the Girardi and Giuricin (\cite{gir:giu}) mass function.
If low surface brightness galaxies were more prevalent in low
mass groups -- which is plausible -- the LCLG mass function would
flatten out as seen in Fig.~\ref{massf}.

In a recent paper Mart{\'\i}nez et al. (\cite{martinez}) found
a similar flattening of the mass function of groups, extracted
from the 2dF catalogue. This is also a deep sample, similar to
the LCLG catalogue.

We should also look for possible problems with our simulated mass
functions.

	One problem is that we are comparing dark matter
halos with galaxy-populated halos. It is probable that
halos of smaller masses ($10^{12}-10^{13}$) can not easily 
host three or more galaxies. This would explain the deficiency
of small-mass galaxy groups. Thus, better simulations of group
catalogues will have to include galaxy formation.

There could also be a problem with mass determination.
We devoted much attention to accurately evaluate the mass estimates 
in simulations; 
in particular, we succeeded in filtering out unbound groups. 
Still, the mass-velocity dispersion relation 
of virialised groups widens toward the small mass end,
showing that virial masses of small groups may have large 
intrinsic errors.  This widening
could be caused, first, by the predominance of radial orbits in small
groups, but this should be a rather small effect. 
Another possibility is
that small groups might have a higher intrinsic velocity
spread than large groups 
due to different formation histories, or in other words, due to a mixture 
of young and old groups. This is a question that could be only answered 
by a detailed dynamical study of observed groups.

Another numerical effect could be an enhanced group production
in simulations, caused by 
gravitational two-body collisions, which were discussed by 
Suisalu \& Saar \cite{suisalu} and Splinter et
al. \cite{splinter}). More recent analysis concerning the convergence of 
dissipationless dark matter $N$-body codes by Knebe et al. (\cite{knebe}) 
and Power et al. (\cite{power}) demonstrates that these effects are small if 
the gravitational softening and the time steps are chosen carefully.

\section{Conclusion}

In this work we have presented the mass function of the Las Campanas Loose 
Groups selected by Tucker et al. (\cite{tuc:tuc}). We have also studied the 
mass-velocity dispersion relation that shows the expected scaling relation 
for virialised systems. We conclude that the loose galaxy groups are 
basically physical systems. The mass function was compared 
with results of dark matter simulations.

Our main conclusions are: 

1) The completeness interval of the LCLG sample is found to be around
$5\cdot10^{13}{\rm M}_{\sun}/h \le M \le 8\times 10^{14}{\rm M}_{\sun}/h$.

2) The high end of the mass function of the LCLG sample 
lies between the simulated mass functions corresponding
to $\sigma_8=0.78$ and $\sigma_8=0.87$.

3) At small masses $M < 5\cdot10^{13}{\rm M}_{\sun}/h$ the 
mass function of the LCRS sample of
loose groups flattens out.

\begin{acknowledgements} 

The present study was supported by the Estonian Science Foundation grant
4695.  P.H. was supported by the Academy of Finland (grant 46733).
D.L.T. was supported by the US Department of Energy under contract 
No. DE-AC02-76CH03000. We thank 
Mirt Gramann, Sahar Allam and Gert H\"utsi for  
comments.  
We acknowledge the use of the AP3M code made
public by H. Couchman. Finally, we thank our referee 
for a number of useful and constructive comments.   

\end{acknowledgements}

\end{document}